\documentstyle[aps,epsf,epsfig]{revtex}
 
\newcommand{\bq}{\begin{equation}}
\newcommand{\eq}{\end{equation}}
\newcommand{\bqn}{\begin{eqnarray}}
\newcommand{\eqn}{\end{eqnarray}}
\newcommand{\nb}{\nonumber}
\newcommand{\lb}{\label}

\def\eax{\frac{(2-\alpha)e^{\frac{\alpha}{2}x}}{2}}
\def\eaxd{\frac{(2-\alpha)e^{\frac{\alpha}{2}x}}{4}}
\def\a{\alpha}
\def\ad{\alpha^2}

\def\pzx{{\Phi_0}^\prime}
\def\pzxx{{\Phi_0}^{\prime \prime}}

\def\pzx{{\Phi_0}^\prime}
\def\pzxx{{\Phi_0}^{\prime \prime}}
\def\pl{\Phi_1}
\def\plx{{\Phi_1}^\prime}
\def\plxx{{\Phi_1}^{\prime \prime}}
\def\psz{\Psi_0}
\def\pszx{{\Psi_0}^\prime}
\def\pszxx{{\Psi_0}^{\prime \prime}}
\def\psl{\Psi_1}
\def\pslx{{\Psi_1}^\prime}
\def\pslxx{{\Psi_1}^{\prime \prime}}
\def\sz{S_0}
\def\szx{{S_0}^\prime}
\def\szxx{{S_0}^{\prime \prime}}
\def\sl{S_1}
\def\slx{{S_1}^\prime}
\def\slxx{{S_1}^{\prime \prime}}

\begin{document}
\title{Perturbed Self-Similar Massless Scalar Field
in the Spacetimes with Circular Symmetry  
 in $2+1$ Gravity}
\author{R. Chan$^1$ \thanks{E-mail Address: chan@on.br},
M.F.A. da Silva $^2$ \thanks{E-mail Address:  mfas@dft.if.uerj.br} and
Jaime F. Villas da Rocha$^2$ \thanks{E-mail Address: roch@dft.if.uerj.br}}
\address{$^2$ Departamento de F\' {\i}sica Te\' orica,
Universidade do Estado do Rio de Janeiro,
Rua S\~ ao Francisco Xavier $524$, Maracan\~ a,
CEP 20550--013, Rio de Janeiro, RJ, Brazil \\
$^1$ Coordenadoria de Astronomia e Astrof\'{\i}sica,
Observat\'orio Nacional, Rua General Jos\'e Cristino 77, S\~ao
Crist\'ov\~ao, CEP 20921--400, Rio de Janeiro, RJ, Brazil}
\date{\today}
\maketitle

\begin{abstract}

We present in this work the study of the linear perturbations 
of the $2+1$-dimensional circularly symmetric solution, obtained 
in a previous work, with kinematic self-similarity of the second kind.
We have obtained an exact solution for the perturbation equations and
the possible perturbation modes.
We have shown that the background solution is a stable solution.

\end{abstract}

\section{Introduction}

One of the most outstanding problems in gravitation theory is the
study of the relation that exists between the critical phenomena
and the process of black hole formation.
The studies of non-linearity of the Einstein field equations near
the threshold
  of
black hole formation reveal very rich phenomena \cite{Chop93},
which are quite similar to critical phenomena in Statistical
Mechanics and Quantum Field Theory \cite{Golden}. In particular,
by numerically studying the gravitational collapse of a massless
scalar field in $3+1$-dimensional spherically symmetric
spacetimes, Choptuik found that the mass of such formed black
holes takes a scaling form,
\bq
\lb{I.1}
M_{BH} = C(p)\left(p -p^{*}\right)^{\gamma},
\eq
where $C(p)$ is a constant and depends on the initial data, and
$p$ parameterises a family of initial data in such a way that when
$p > p^{*}$  black holes are formed, and when $p < p^{*}$ no black
holes are formed. It was shown that, in contrast to $C(p)$, the
exponent $\gamma$ is universal to all the families of initial
data studied. Numerically it was determined as $\gamma \sim 0.37$.
The solution with $p = p^{*}$, usually called the critical
solution, is found also universal. Moreover, for the massless
scalar field it is periodic, too. Universality  of the critical
solution and exponent, as well as the power-law scaling of the
black hole mass all have given rise to the name {\em Critical
Phenomena in Gravitational Collapse}. Choptuik's studies were soon
generalised to other matter fields \cite{Gun00,Wang01}, and now
the following seems clear: (a) There are two types of critical
collapse, depending on whether the black hole mass takes the
scaling form (\ref{I.1}) or not. When it takes the scaling form,
the corresponding collapse is called Type $II$ collapse, and when
it does not it is called Type $I$ collapse. In the type $II$
collapse, all the critical solutions found so far have either
discrete self-similarity (DSS) or homothetic self-similarity
(HSS), depending on the matter fields. In the type $I$ collapse,
the critical solutions have neither DSS nor HSS. For certain
matter fields, these two types of collapse can co-exist. (b) For
Type $II$ collapse, the corresponding exponent is universal only
with respect to certain matter fields. Usually, different matter
fields have different critical solutions and, in the sequel,
different exponents. But for a given matter field the critical
solution and the exponent are universal.  So far, the
studies have been mainly restricted to spherically symmetric case
and their non-spherical linear perturbations. Therefore, it is not
really clear whether or not the critical solution and exponent are
universal with respect to different symmetries of the
spacetimes \cite{Cho03,Wang03}. (c)  A critical solution for both of the two
types has one and only one unstable mode. This now is considered as one
of the main criteria for a solution to be critical. (d) The
universality of the exponent is closely  related to the last
property. In fact, using dimensional analysis \cite{Even} one can
show that
\bq
\lb{I.2}
\gamma = \frac{1}{\left|k\right|},
\eq
where $k$ denotes the unstable mode.

From the above, one can see that to study (Type $II$)
critical collapse, one
may first find some particular solutions by imposing certain
symmetries, such as, DSS or HSS. Usually this considerably
simplifies the problem. For example, in the spherically symmetric
case, by imposing HSS symmetry the Einstein field equations can be
reduced from PDE's to ODE's.    Once the particular solutions are
known, one can study their linear perturbations and find out the
spectrum of the corresponding eigenmodes. If a solution has one
and only one unstable mode, by definition we may consider it as a
critical solution (See also the discussions given in
\cite{Brady02}). The studies of critical collapse  have been
mainly numerical so far, and analytical ones are still highly
hindered by the complexity of the problem, even after imposing
some symmetries.

Lately, Pretorius and Choptuik (PC) \cite{PC00} studied
gravitational collapse of a massless scalar field in an anti-de
Sitter background in $2+1$-dimensional spacetimes with circular
symmetry, and found that the  collapse exhibits  critical
phenomena and the mass of such formed black holes takes the
scaling form of Eq.(\ref{I.1}) with $\gamma = 1.2 \pm 0.02$, which
is different from that of the corresponding $3+1$-dimensional
case.  In addition, the critical solution is also different, and,
instead of having DSS, now has HSS. The above results were
confirmed by independent numerical studies \cite{HO01}. However,
the exponent obtained by Husain and Olivier (HO),  $\gamma \sim
0.81$,  is quite different from the one obtained by  PC. It is not
clear whether the difference is due to numerical errors or to some
unknown physics.

After the above numerical work, analytical studies of the same
problem soon followed up \cite{Gar01,CF01,GG02,HWW04}. In
particular, Garfinkle found a class, say, $S[n]$, of exact
solutions to the Einstein-massless-scalar field equations and
showed that in the strong field regime the $n = 4$ solution fits
very well with the numerical critical solution found by PC.
Lately, Garfinkle and Gundlach (GG) studied their linear
perturbations and found that only the solution with $n = 2$ has
one unstable mode, while the one with $n = 4$ has three
\cite{GG02}. According to  Eq.(\ref{I.2}), the corresponding
exponent is given by $\gamma = 1/|k| = 4/3$. Independently,
Hirschmann, Wu and one of the present authors (HWW) systematically
studied the problem, and  found that the $n = 4$ solution indeed
has only one unstable mode \cite{HWW04}. This difference actually
comes from the use of different boundary conditions. As a matter
of fact, in addition to the ones imposed by GG \cite{GG02}, HWW
further required that no matter field should come out of the
already formed black holes. This additional condition seems
physically quite reasonable and has been widely used in the
studies of black hole perturbations \cite{Chandra83}. However, now
the corresponding exponent is given by $\gamma = 1/|k| = 4$,
which is significantly different from the numerical ones. So far,
no  explanations about these differences have been worked out,
yet.

Self-similarity is usually
divided into two classes, one is the discrete self-similarity
mentioned above, and the other is the so-called kinematic
self-similarity (KSS) \cite{CH89}, and sometimes it is also called
continuous self-similarity (CSS). KSS or CSS is further classified
into three different kinds, the zeroth, first and second. The
kinematic self-similarity of the first kind is also called
homothetic self-similarity, first introduced to General Relativity
by Cahill and Taub in 1971 \cite{CT71}. In Statistical Mechanics,
critical solutions with KSS of the second kind seem more generic
than those of the first kind \cite{Golden}. However,  critical
solutions with KSS of the second kind have not been found so far
in gravitational collapse, and it would be very interesting to
look for such solutions.   
We shall present in this work the study of the linear perturbations 
of the $2+1$-dimensional circularly symmetric solution, obtained 
in a previous work \cite{CFJW04}, with kinematic self-similarity of 
the second kind and show that the background solution is not critical.

In Section II we present the field equations with kinematic self-similarity 
of the second kind.  In Section III we perturb linearly the field equations.
In Section IV we present the solution of the linear perturbation equations.
I Section V we apply the boundary conditions in the perturbed solutions and
in Section VI we conclude our work.

\section{The Field Equations with Kinematic Self-similarity of the
Second Kind}

The general metric for such spacetimes can be written in the form
\bq
\lb{1.1}
ds^{2} = e^{2\Phi(r,t)}dt^2-e^{2\Psi(r,t)}dr^2-r^{2}S(r,t)^2 d\theta^2.
\eq
Then, the corresponding non-vanishing components of the Ricci tensor
are
\bqn
\lb{1.2a}
R_{tt} &=& e^{2(\Phi-\Psi)}\left[\Phi_r\left(\Phi_r-\Psi_r+\frac{S_r}{S}
+\frac{1}{r}\right)+\Phi_{rr}\right]-\frac{S_{tt}}{S}+\Phi_t
\frac{S_t}{S}+\Phi_t\Psi_t-{\Psi_t}^2-\Psi_{tt}\nb\\
R_{tr} &=&
\frac{\Psi_t}{r}+\Psi_t\frac{S_r}{S}+\Phi_r\frac{S_t}{S}-\frac{S_t}{rS}-
\frac{S_{rt}}{S}\nb\\
R_{rr} &=&
e^{2(\Psi-\Phi)}
\left[\Psi_{tt}+\Psi_t\left({\Psi_t}+\frac{S_t}{S}-\Phi_t\right)\right]
-\Phi_{rr}+\Phi_r\Psi_r-{\Phi_r}^2-\frac{S_{rr}}{S}-2\frac{S_r}{rS}
+\Psi_r\frac{S_r}{S}+\frac{\Psi_r}{r}\nb\\
R_{\theta\theta} &=&
r^2 S^2\left\{e^{-2\Phi}\left[\frac{S_t}{S}
\left(\Psi_t-\Phi_t\right)+\frac{S_{tt}}{S}\right]-e^{-2\Psi}\left[\frac{S_r}{S}\left(\frac{2}{r}
+\Phi_r-\Psi_r\right)+\frac{1}{r}(\Phi_r-\Psi_r)+\frac{S_{rr}}{S}\right]\right\},
\eqn
where the indices $t$ and $r$ denote differentiation to the time
coordinate and
the radial coordinate, respectively.

Then, introduce two self-similar variables $\tau$ and $x$ via the
relations \bq \lb{1.3a} x= \ln \left(\frac r{\left(-t\right)^{\frac 1\alpha
}}\right),\;\;\;\;\; \tau =-\ln \left(-t\right),
\eq
or inversely,
\bq
\lb{1.3b}
r = e^{(\alpha x - \tau)/\alpha},\;\;\;
t = - e^{-\tau},
\eq
where $\alpha$ is a {\it dimensionless} constant.  
For any given function $f\left(t,r\right) $, we have
\bqn
\lb{1.4}
f_{,t} &=&-\frac 1{\alpha t}\left(\alpha
f_{,\tau }+f_{,x}\right),\;\;\;\;
f_{,r}=\frac 1rf_{,x},  \nonumber \\
f_{,tr} &=&-\frac 1{\alpha tr}\left( \alpha f_{,\tau
x}+f_{,xx}\right),\;\;\;\;\;
f_{,rr}=\frac 1{r^2}\left( f_{,xx}-f_{,x}\right) ,  \nonumber \\
f_{,tt} &=&\frac 1{\alpha ^2t^2}\left( \alpha ^2f_{,\tau \tau }
+2\alpha f_{,\tau x}+f_{,xx}+\alpha ^2f_{,\tau }+\alpha
f_{,x}\right),
\eqn
where the comma means differentiation.

Substituting these equations into Eq.(\ref{1.2a}), we find that
in terms of the self-similar variables, the Ricci tensor is given by
  
\bqn
\lb{1.2b}
R_{tt} &=&
\frac{e^{2(\Phi-\Psi)}}{r^2}\left\{\Phi_x(\Phi_x-\Psi_x+\frac{S_x}{S})+
\Phi_{xx}\right\}-\frac{1}{\alpha^2
t^2}\left\{\alpha^2\left[\Psi_{\tau\tau}+
\Psi_\tau(1+\Psi_\tau-\Phi_\tau)+\frac{S_{\tau\tau}}{S}+\frac{S_\tau}{S}
(1-\Phi_\tau)\right]\right. \nb\\
&+&\alpha\left[2\Psi_{\tau x}+\Psi_\tau(\Psi_x-\Phi_x)+
\Psi_x(\Psi_\tau-\Phi_\tau+1)+2\frac{S_{\tau
x}}{S}+\frac{S_x}{S}(1-\Phi_\tau)-\frac{S_\tau}{S}\Phi_x\right] \nb \\
&+&\left. \left[\Psi_{xx}+
\Psi_x(\Psi_x-\Phi_x)+\frac{S_{xx}}{S}-\frac{S_x}{S}\Phi_x\right]\right\}
\nb\\
R_{tr} &=& -\frac{1}{\alpha
tr}\left\{\alpha\left[\Psi_\tau\left(1+\frac{S_x}{S}\right)+\frac{S_\tau}{S}
\left(\Phi_x-1\right)-\frac{S_{\tau x}}{S}\right]+\Psi_x
\left(1+\frac{S_x}{S}\right)
+\frac{S_x}{S}\left(\Phi_x-1\right)-\frac{S_{xx}}{S}\right\}\nb\\  
R_{rr} &=&\frac{e^{2(\Psi-\Phi)}}{\alpha^2
t^2}\left\{\alpha^2\left[\Psi_{\tau\tau}+\Psi_\tau\left(1+\Psi_\tau+
\frac{S_\tau}{S}-\Phi_\tau\right)\right]\right.\\ \nb
& & \left.+\alpha\left[2\Psi_{\tau
x}+\Psi_x\left(1+\Psi_\tau+\frac{S_\tau}{S}-\Phi_\tau\right)+
\Psi_\tau\left(\Psi_x+\frac{S_x}{S}-\Phi_x\right)\right]\right.\nb\\
&+&\left.\Psi_{xx}+\Psi_x\left(\Psi_x+\frac{S_x}{S}-\Phi_x\right)\right\}+
\frac{1}{r^2}\left[\Phi_x\left(\Psi_x-\Phi_x+1\right)-\Phi_{xx}+
\Psi_x\left(1+\frac{S_x}{S}\right)-\frac{1}{S}\left(S_{xx}+S_x\right)\right]
\nb\\
R_{\theta\theta} &=& r^2S^2\left\{\frac{e^{-2\Phi}}{\alpha^2 t^2
S}\left[\alpha^2\left(S_\tau\left(1+\Psi_\tau-\Phi_\tau\right)+S_{\tau\tau}\right)+
\alpha\left(S_\tau\left(\Psi_x-\Phi_x\right)+S_x\left(1+\Psi_\tau-\Phi_\tau\right)+
2S_{\tau x}\right)\right.\right.\nb\\
&+&\left.\left.S_x\left(\Psi_x-\Phi_x\right)+S_{xx}\right]
-\frac{e^{-2\Psi}}{r^2}\left[\frac{1}{S}\left(S_x\left(1+\Phi_x-\Psi_x\right)+
S_{xx}\right)+\Phi_x-\Psi_x\right]\right\},
\eqn
where the indices $\tau$ and $x$ mean differentiation in respect to
these variables.

\section{The Linear Perturbation of the Field Equations}

Once we have the general expressions of $R_{\mu\nu}$ in terms of $\tau$ and
$x$, let us consider the following perturbations,
\bqn
\lb{1.5a}
\Phi(\tau, x) &=& \Phi_{0}(x) + \epsilon \Phi_{1}(x)e^{k\tau},\nb\\
\Psi(\tau, x) &=& \Psi_{0}(x) + \epsilon \Psi_{1}(x)e^{k\tau},\nb\\
S(\tau, x) &=& S_{0}(x) + \epsilon  S_{1}(x)e^{k\tau},\nb\\
\lb{1.5}
\phi(\tau, x) &=& \phi_{0}(t) + \epsilon \phi_{1}(x)e^{k\tau},
\eqn
where $\epsilon$ is a very small real constant, the quantities with
subscripts
``1" denote perturbations, those with ``0" denote the background 
self-similar solutions.

The background solution is given by
\bqn
\lb{backsol}
\Phi_0(x) &=& 0, \nb\\
\Psi_0(x) &=& -\frac{1}{2}\alpha x, \nb\\ 
S_0(x) &=& \frac{2}{2-\alpha}e^{-\frac{1}{2}\alpha x},\nb\\
\phi_{0}(t) &=& 2q\ln(-t),
\eqn
and the apparent horizon is given by
\bq
\lb{rAHt}
r_{AH}(t) = \left[\left(2-\alpha\right)(-t)^{1/2}\right]^{2/(2-\alpha)}.
\eq
where $\varphi_0$, $\psi_{0}$ and $s_{0}$ are integration constants,
$q = \pm \frac{1}{\sqrt{8}}$ and $\alpha < 2$ \cite{CFJW04}.

It is understood that there may be many perturbation modes for different
values (possibly complex) of the constant $k$.  The general perturbation
will be the sum of these individual modes.  Those modes with $Re(k) > 0$
grow as $\tau \rightarrow
\infty$ and are referred to as unstable modes, while the ones with
$Re(k) < 0$ decay and are referred to as stable modes.  By definition,
critical solutions will have one and only one unstable mode.  

Substituting Eq.(\ref{1.5}) into Eq.(\ref{1.2b}) and then we will have
\bq
\lb{1.7}
R_{\mu\nu} = R_{\mu\nu}(\tau, x, \epsilon).
\eq
Now considering $R_{\mu\nu}$ is function of $\epsilon$ only, then we
expand it
in terms of $\epsilon$,
\bq
\lb{1.8}
R_{\mu\nu}(\tau, x, \epsilon) =
\frac{1}{(-t)^{2}}\left\{R_{\mu\nu}^{(0)}(x)
+ \epsilon R_{\mu\nu}^{(1)}(x)e^{k\tau} +
O\left(\epsilon^{2}\right) \right\},
\eq
where $R_{\mu\nu}^{(0)}(x)$ is the part of the Ricci tensor corresponding
background,
and $R_{\mu\nu}^{(1)}(x)$ the perturbation part, which is function of the
background, $\Phi_{0}(x), \Psi_{0}, S_{0}(x)$ and the linear perturbations,
$\Phi_{1}(x), \Psi_{1}, S_{1}(x)$.
In the paper of Hirschmann, Wang \& Wu \cite{HWW04} it was calculated
them for the self-similar solutions of the first kind but in double null
coordinates [cf. Eq.(65) given there].  

To first order in $\epsilon$, it can be shown that the
non-vanishing components of the Ricci tensor are given by
\bqn
\lb{eqsa}
R_{tt}^{(1)}(x)&=&
e^{2\left(\Phi_0 - \Psi_0 -x +\frac{\tau}{\alpha} \right)}
\left\{ \pzx \left( 2\plx - \pslx \right) - \plx \pszx +\plxx
+ 2 \left( \pl - \psl \right)\left[ \pzx \left(\pzx -\pszx\right)
+ \pzxx \right] {\phantom{\frac{1}{1}}}   \right.  
\nb \\  & & \left.
+ \frac{1}{\sz}\left[
-  \frac{\szx \sl}{\sz} \pzx
+ 2 \left( \pl - \psl \right) \pzx \szx + \pzx \slx  +\plx \szx \right]
 \right\} +
\\ \nb & & +
\frac{e^{2 \tau}}{\alpha^2}
\left\{  \pzx \left(
\alpha k \psl +\pslx \right) +
\pszx \left(\alpha k \pl +\plx \right) -
2 \pszx\left(\alpha k \psl +\pslx \right) -
\alpha^2 k^2 \psl - 2 \alpha k \pslx - \pslxx
\right. \\ \nb &  & {\phantom{
\frac{e^{2 \tau}}{\alpha^2}}}
- \alpha^2 k \psl
- \alpha \pslx  +\\ \nb & &
+ \frac{1}{S_0} \left[
\pzx \left( \alpha k S_1 + \slx \right)
+ \szx \left( \alpha k \pl + \plx \right)
- \alpha^2 k^2 S_1 - 2 \alpha k \slx - \slxx -
\alpha^2 k S_1 - \alpha \slx \right. \\ \nb &&
{\phantom{
\frac{e^{2 \tau}}{\alpha^2}}} \left. \left.
- \frac{1}{S_0} \left(
\szx S_1  \pzx - \szxx S_1 - \alpha \szx S_1
\right) \right] \right\} \\
R_{tr}^{(1)}(x)&=&
-\frac{e^{\frac{\alpha +1}{\alpha}\tau - x}}{\alpha S_0}
\left[ - \frac{S_1}{S_0} \left( \szx - \pzx \szx -\pszx \szx + \szxx -
\sz \pszx
\right) - \pzx \left( \alpha k S_1 + \slx \right) - \pszx \slx
\right. +
\nb\\
& & \phantom{{-\frac{e^{\frac{\alpha +1}{\alpha}\tau - x}}{\alpha}}}
\left. -
\szx \left( \alpha k \psl + \pslx + \plx \right)
+ \alpha k \slx + \slxx - \pszx S_1
- S_0 \left( \alpha k \psl + \pslx  \right) +
\alpha k \sl + \slx  
\right]
\\ \nb
R_{rr}^{(1)}(x) &=&
\frac{e^{2 \left(\Psi_0 -\Phi_0 + \tau \right)}}{\alpha^2}
\left[ 2\left(\psl - \pl \right)
\left( {\pszx}^2+\pszxx + \alpha \pszx - \pzx \pszx + \pszx \szx  \right)
+ 2 \alpha k \psl \pszx + 2 \pszx \pslx + \alpha^2 k^2 \psl
\right. +
\nb \\ & &
{\phantom{\frac{e^{2\left(\Psi_0 - \Phi_0 + \tau \right)}}{\alpha^2}}}
+ 2 \alpha k \pslx + \pslxx + \alpha^2 k \psl + \alpha \pslx
- \alpha k \psl \pzx - \pzx \pslx - \alpha k \pszx \pl
- \pszx \plx + \nb
\\ & & \phantom{{-\frac{e^{\frac{\alpha +1}{\alpha}\tau - x}}{\alpha}}}
\left.
\frac{1}{S_0^2}
\left( \alpha k S_0 \pszx S_1 + \sz \pszx \slx + \alpha k
\sz \szx \psl + \sz \szx \pslx - \sl \pszx \szx \right) \right] +
\nb\\ & & -
e^{2\left( \frac{ \tau}{\alpha} -x \right)}
\left[
- \plx \pszx + \plxx - \plx + 2 \pzx \plx - \pzx
\pslx - \frac{1}{S_0} \left(
\pszx \slx + \pslx \szx + \sz \pslx - \slxx - \slx \right) \right. +
\nb \\ && {\phantom{e^{2\left( \frac{ \tau}{\alpha} -x \right)}}}
+ \left. \frac{S_1}{S_0^2} \left(
\pszx \szx - \szxx -\szx \right) \right]
\nb  \\
R_{\theta\theta}^{(1)}(x) &=&
-e^{-2\psz} \left[ \left( \sl - 2\sz \psl\right)
\left( \pzx \szx + \pzx \sz - \pszx \szx - \pszx \sz + \szxx +\szx \right)
\right. +
\nb
\\ && \left.
+ \sz \left( \pzx \slx + \pzx \sl + \plx \szx + \plx \sz
- \pszx \slx - \pszx \sl - \pslx \szx - \pslx \sz + \slxx + \slx \right)
\right] -
\\ && -
\frac{e^{2 \left(\frac{\alpha-1}{\alpha} \tau +x -\Phi_0
 \right)}}{\alpha^2}
\left[
\left( \sl - 2\sz \pl\right)
\left( \pzx \szx - \pszx \szx - \szxx -\alpha \szx\right)
\phantom{\frac{}{1}} \right. +
\nb \\ && +
\sz \left( \alpha k \pzx \sl + \pzx \slx
+ \alpha k \pl \szx + \plx \szx - \alpha k \sl \pszx
-\pszx \slx - \alpha k \psl \szx +
\right.
\nb \\ &&
\left. \left.
- \pslx \szx - \alpha^2 k^2 \sl - 2\alpha k
\slx - \slxx - \alpha^2 k \sl - \alpha \slx \right) \right],
\nb
\eqn
where the prime denotes differentiation in respect to $x$.

Once we have $R_{\mu\nu}^{(1)}(x)$, we have to calculate the quantities
\bq
\lb{1.9}
A_{\mu\nu} \equiv \phi_{,\mu}\phi_{,\nu}.
\eq
Substituting Eqs.(\ref{1.5}) into the above equations, we have
\bq
\lb{1.10}
A_{\mu\nu}(\tau, x, \epsilon) =
\frac{1}{(-t)^{2}}\left\{A_{\mu\nu}^{(0)}(x)
+ \epsilon A_{\mu\nu}^{(1)}(x)e^{k\tau} +
O\left(\epsilon^{2}\right) \right\},
\eq
where  $A_{\mu\nu}^{(0)}(x)$ is the part of the  background,
and $A_{\mu\nu}^{(1)}(x)$ the perturbation part, which is function of the
background, $\phi_{0}(t)$, and the linear perturbation,
$\phi_{1}(x)$, and given by
\bqn
\lb{eqsb}
A_{tt}^{(1)}(x)&=& -\frac{e^{2\tau}}{\alpha} \left[4 q \left(
\alpha k \phi_1 + \phi_1^\prime \right) \right]\nb\\
A_{tr}^{(1)}(x)&=&
- \frac{e^{\frac{\alpha +1}{\alpha}\tau-x}}{\alpha}
\left[ 2 q \phi_1^\prime \right]
\nb\\
A_{rr}^{(1)}(x) &=& 0
\nb\\
A_{\theta\theta}^{(1)}(x) &=& 0,
\eqn
where the dot means time differentiation.

Once we have $A_{\mu\nu}^{(1)}(x)$ and $R_{\mu\nu}^{(1)}(x)$, the linear
perturbation equations are given by
\bq
\lb{eqs}
R_{\mu\nu}^{(1)}(x) = A_{\mu\nu}^{(1)}(x),
\eq
which in general are complicated.

After we have the general linear perturbation
equations (\ref{eqs}), then we turn to consider the background solutions
given by Eqs.(\ref{backsol}). By virtue of the simple form of
the solutions and the fact $\phi_{0}(x) = 0$, Eqs.(\ref{eqs}) can be 
solved in our case.

\bqn
\lb{sisteq1}
\ad  k \pl + \a \plx +\ad k^2 \psl + 2\a k \pslx + \pslxx
+\eaxd \left[ \left( 2 k^2 +2 k +\frac{1}{2} \right) \ad \sl
+ 2 \a \left( 2 k +1 \right) \slx + 2 \slxx \right]  
= & & \\  \nb
-4 q \a \left(\a k \phi_1 + \phi_1^\prime \right), & &
\eqn

\bq
\lb{sisteq2}
\plxx= 0,
\eq
\bq
\lb{sisteq3}
(\a -2)\left[\a k \psl + \pslx \right] +
\a \plx + \eax \left[ \a \left( 2 k + 1 \right) \sl + \left( 2 \a k +
\a +2 \right) \slx + 2 \slxx \right]
= - 2 q \phi_1^\prime,
\eq
\bq
\lb{sisteq4}
\ad k \left( 2 k -1 \right) \psl +
\a \left( 4 k -1\right) \pslx
+ 2 \pslxx + \a \left( \a k \pl + \plx \right)
- \a \eaxd \left[ \left( 2 \a k + \a \right) \sl + 2 \slx \right]
 = 0,
\eq
\bq
\lb{sisteq5}
\left( 2 - \a \right) \pslx + \left( 2  - \a \right) \plx - 2 \plxx
- \eax \left[ \a \sl + \left( 2 + \a \right) \slx + 2 \slxx \right]
= 0,
\eq
\bq
\lb{sisteq6}
\a \left( \a k \psl + \pslx \right)
- \a \left( \a k \pl + \plx \right) -
\eax \left[ \ad k \left( 1 + 2 k \right) \sl + \a \left( 1 + 4 k \right)
\slx + 2 \slxx \right]
= 0,
\eq
\bq
\lb{sisteq7}
\left( 2 - \a \right) \pslx - \left( 2  - \a \right) \plx
- \eax \left[ \a \sl + \left( 2 + \a \right) \slx + 2 \slxx \right]
= 0.
\eq                                                          

\section{The Solutions of the Linear Perturbation Equations}

We will solve the system of the perturbed Eqs.(\ref{sisteq1})-(\ref{sisteq7}).
From Eq.(\ref{sisteq2}) we have
\bq
\lb{phi1}
\Phi_1=ax+b.
\eq

From Eqs.(\ref{sisteq5}) and (\ref{sisteq7}) we have
\bq
\lb{sisteq5-7}
(2-\alpha)\Phi_1'=0,
\eq
which solutions are $\alpha=2$ (which is out of range of our solution)
or $\Phi_1'=0$.  Thus, from Eq.(\ref{phi1}) we have that
\bq
\lb{phi1xb}
\Phi_1=b=constant.
\eq

Using Eq.(\ref{phi1xb}) and summing the Eq.(\ref{sisteq5})
and Eq.(\ref{sisteq7}) we get
\bq
\lb{sisteq5+7}
\Psi_1'={{e^{{1 \over 2} \alpha x}} \over {2}}[\alpha S_1 +
(2+\alpha)S_1'+ 2S_1'']
\eq

Using Eq.(\ref{phi1xb}) and substituting Eq.(\ref{sisteq5+7}) into 
Eq.(\ref{sisteq6}) we get
\bq
\lb{psi1x}
\Psi_1=b-{\Psi_1' \over {\alpha k}}+{{(2-\alpha)e^{{1 \over 2} \alpha x}} \over
{2\alpha^2k}}[\alpha^2k(1+2k)
S_1 + \alpha(1+4k)S_1'+ 2S_1'']
\eq

Substituting Eq.(\ref{sisteq5+7}) into Eq.(\ref{psi1x}) and
differentiating it, we have
\bq
\lb{dfeqS1x}
A S_1 + B S_1' + C S_1'' + 4 S_1''' + E e^{-{1 \over 2}\alpha x} = 0,
\eq
where
\bq
\lb{A}
A = {\alpha^2 \over 2} (- 8\alpha k^3 + 4 \alpha k + \alpha + 16 k^3 - 4 k),
\eq
\bq
\lb{B}
B =  \alpha ( - 8\alpha k^2  + 4\alpha k + 3\alpha + 16 k^2 ),
\eq
\bq
\lb{C}
C = 2 (8\alpha k -\alpha - 4 k + 4),
\eq
\bq
\lb{E}
E =  4\alpha^2  b k^2.
\eq

Since our background solution with second kind self-similarity is identical 
to the solution with self-similarity of the first kind \cite{CFJW04}, we will
study hereinafter only the case $\alpha=1$.  Thus,
\bq
A = 4k^3+\frac{1}{2},
\eq
\bq
B = 8 k^2  + 4 k + 3,
\eq
\bq
C = 8 k + 6,
\eq
\bq
E = 4bk^2,
\eq
and the solution of equation (\ref{dfeqS1x}) is given by
\bq
S_1(x) = -\frac{b e^{-\frac{1}{2} x}}{k-1}+c_1 e^{-\frac{1}{2} (1+2 k) x}+c_2 e^{-\frac{1}{2} (k+1+\sqrt{\Delta}) x}+c_3 e^{-\frac{1}{2} (k+1-\sqrt{\Delta}) x}.
\eq
where
\bq
\Delta= -k (3 k-4).
\eq

In the next section we will apply the boundary conditions for two special
cases: $\Delta > 0$ and $\Delta < 0$.


\section{The Boundary Conditions for the Perturbed Solutions}

We will apply the boundary conditions only at two regions of the spacetime:
at the centre of the spacetime $r = 0$  and 
at the event horizon $r_{AH}$ given by equation (\ref{rAHt}), 
that furnishes
\bq
\lb{xAH}
r_{AH}=-t.
\eq
Thus, the metric at the apparent horizon is given by
\bq
\lb{ds2AH}
ds^2_{AH}=dt^2 - dr^2 - 4 (-t)^2 d \theta^2.
\eq
It can be easily seen from this metric that the apparent horizon is singular,
in this case, only at $t$ $=$ 0. Then the final state of the collapse is a marginally naked singularity.

We would like to note that for the perturbed part of the metric (\ref{1.1})
to represent circular symmetry, some physical and geometrical
conditions needed to be imposed \cite{Yasuda}. For gravitational
collapse, we impose the following conditions at $r=0$ :
 
\begin{description}

\item ($i$) There must exist a symmetry axis, which can be expressed as
\bq
\lb{cd1}
X \equiv \left|\xi^{\mu}_{(\theta)}\xi^{\nu}_{(\theta)}
g_{\mu\nu} \right| \rightarrow 0,
\eq
as $r \rightarrow 0$ , we have chosen the radial coordinate
such that the axis is located at $r = 0$ , and
$\xi^{\mu}_{(\theta)}$ is the Killing vector with a close orbit,
and given by $\xi^{\alpha}_{(\theta)}\partial_{\alpha} =
\partial_{\theta}$.
 
\item ($ii$) The spacetime near the symmetry axis is locally flat, which
can be written as \cite{Kramer80}
\bq
\lb{cd2}
\frac{X_{,\alpha}X_{,\beta} g^{\alpha\beta}}{4X}
\rightarrow - 1,
\eq
as $r \rightarrow 0$ . 
Note that solutions failing to satisfy this
condition sometimes are also acceptable. For example, when the
left-hand side of the above equation approaches a finite constant,
the singularity at $r = 0$  
may be related to a point-like particle \cite{VS}. 
 
\item ($iii$) No closed timelike curves (CTC's). In spacetimes with
circular symmetry, CTC's can be easily introduced. To ensure
their  absence, we assume that the condition
\bq
\lb{cd3}
\xi^{\mu}_{(\theta)}\xi^{\nu}_{(\theta)}g_{\mu\nu} < 0,
\eq
holds in the whole spacetime.

\end{description} 

\subsection{Case $\Delta > 0$}

In this case we have from equation (\ref{psi1x}) that
\bqn
\Psi_1(x)&=& \frac{1}{k-1} \left[ k^2 c_3 e^{\frac{1}{2} x (k+\sqrt{-k (3 k-4)})}+k^2 c_2 e^{\frac{1}{2} x (k-\sqrt{-k (3 k-4)})}+3 e^{x k} k b-2 c_2 k e^{\frac{1}{2} x (k-\sqrt{-k (3 k-4)})} \right. \nonumber \\
&+& \left. c_3 \sqrt{-k (3 k-4)} k e^{\frac{1}{2} x (k+\sqrt{-k (3 k-4)})}-2 c_3 k e^{\frac{1}{2} x (k+\sqrt{-k (3 k-4)})}-c_2 \sqrt{-k (3 k-4)} k e^{\frac{1}{2} x (k-\sqrt{-k (3 k-4)})} \right. \nonumber \\
&-&\left. c_3 \sqrt{-k (3 k-4)} e^{\frac{1}{2} x (k+\sqrt{-k (3 k-4)})}-2 b e^{x k}+c_2 e^{\frac{1}{2} x (k-\sqrt{-k (3 k-4)})}+c_2 \sqrt{-k (3 k-4)} e^{\frac{1}{2} x (k-\sqrt{-k (3 k-4)})} \right. \nonumber \\
&+& \left. c_3 e^{\frac{1}{2} x (k+\sqrt{-k (3 k-4)})} \right] e^{-x k}+c_4 e^{-x k}.
\eqn

In order to apply the first boundary condition (\ref{cd1}), we have to calculate the quantity 
$\sqrt{X}=r S_1$, which can be written as
\bq
rS_1=\frac{b}{k-1}(-r t)^{\frac{1}{2}}+c_1 r^{\frac{1}{2}-k}(-t)^{\frac{1}{2}+k} +
c_2 r^{-\frac{1}{2}(k-1+\sqrt{\Delta})}(-t)^{\frac{1}{2}(k+1+\sqrt{\Delta})}+
c_3 r^{-\frac{1}{2}(k-1-\sqrt{\Delta})}(-t)^{\frac{1}{2}(k+1-\sqrt{\Delta})}.
\eq
Since the limit of $rS_1$ must vanishes when $r \rightarrow 0$, all 
the exponents
of $r$ must be greater than zero.  It is easily shown that some the exponents
cannot satisfy this condition, then the first condition is not fulfilled.
Thus, these perturbations are limited by the boundary conditions.

\subsection{Case $\Delta < 0$}

In this case we have from equation (\ref{psi1x}) that
\bqn
\Psi_1(x)&=&-\frac{e^{-\frac{1}{2} x k}}{k-1} \left[ -2 c_2 k^2 \cos \left(\frac{1}{2} \sqrt{k (3 k-4)} x \right)+4 c_2 k \cos \left( \frac{1}{2} \sqrt{k (3 k-4)} x \right)-3 k b e^{\frac{1}{2} x k} \right. \nonumber \\
&+& \left. 2 c_2 k \sqrt{k (3 k-4)} \sin \left( \frac{1}{2} \sqrt{k (3 k-4)} x \right)-2 c_2 \sqrt{k (3 k-4)} \sin \left( \frac{1}{2} \sqrt{k (3 k-4)} x \right)+2 b e^{\frac{1}{2} x k} \right. \nonumber \\
&-& \left. 2 c_2 \cos \left( \frac{1}{2} \sqrt{k (3 k-4)} x \right) \right]+c_4 e^{-x k}.
\eqn

Then we have now two possibilities for the arbitrary constants $c_2$ and
$c_3$ in order to get a real function $S_1(x)$: $c_2 = c_3$ and $c_2 = -c_3$.

For $c_2 = c_3$ we get 
\bq
S_1(x) = -\frac{b e^{-\frac{1}{2} x}}{k-1}+c_1 e^{-\frac{1}{2} (1+2 k) x}+2c_2 e^{-\frac{1}{2} (k+1+) x} \cos \left( \frac{1}{2}\sqrt{-\Delta} x \right),
\eq
and $rS_1$ is given by
\bq
rS_1=\frac{b}{k-1}(-r t)^{\frac{1}{2}}+c_1 r^{\frac{1}{2}-k}(-t)^{\frac{1}{2}+k} +
2c_2 r^{-\frac{1}{2}(k-1)}\cos\left[ \frac{1}{2}\sqrt{-\Delta} \ln \left( \frac{r}{-t} \right) \right](-t)^{\frac{1}{2}(k+1)}.
\eq

Applying again 
the condition (\ref{cd1}), 
we can see that all the exponents of $r$ must be
greater than zero only when $k < 0$, which admits only stable modes for the
perturbation. 
For the second boundary
condition (\ref{cd2}), 
we have surveyed several sets of values of $b$, $c_1$,
$c_2$, $c_4$ and $k$, and we have found at least one set that satisfies this
condition: $b=0$, $c_1=0$, $c_2=1$, $c_4=1$, $k=-1$.  In this case we get 
\bq
lim_{r \rightarrow 0} \; rS_1 = -7-4\sqrt{7},
\eq
or
\bq
lim_{r \rightarrow 0} \; rS_1 = -4+4\sqrt{7}.
\eq

For $c_2 = -c_3$ we get 
\bq
S_1(x) = -\frac{b e^{-\frac{1}{2} x}}{k-1}+c_1 e^{-\frac{1}{2} (1+2 k) x}-2 i c_2 e^{-\frac{1}{2} (k+1+) x} \sin \left( \frac{1}{2}\sqrt{-\Delta} x \right).
\eq
Since 
this case is analogous to the case where $c_2=c_3$, we will not present it.

\section{Conclusions}

We have presented in this work the study of the linear perturbations
of the $2+1$-dimensional circularly symmetric solution, obtained
by Chan, da Silva, Villas da Rocha \& Wang 
\cite{CFJW04}, with kinematic 
self-similarity of the second kind.  We have shown there that the
solution is Kantowski-Sachs like \cite{Kramer80} and it may be considered
as representing a Friedmann-like cosmological model in a 2+1 dimensional 
spacetime.
We have obtained in this paper an exact solution for the perturbed 
equations, which admits only stable modes,
showing that our Friedmann-like background solution in 2+1 dimension is stable.
This result is in agreement with the conclusions of the work of
Hirschmann, Wang \& Wu \cite{HWW04}. 

\section*{Acknowledgments}

The authors would like to thank Dr. Anzhong Wang for the helpful discussions
and suggestions. The financial assistance from UERJ (JFVdaR) and FAPERJ/UERJ
(MFAdaS) is gratefully acknowledged. The author (R.C.) acknowledges the
financial  support from FAPERJ (no. E-26/171.754/2000 and E-26/171.533/2002).



\begin{thebibliography}{100}

\bibitem{Chop93} M. W. Choptuik, Phys. Rev. Lett. {\bf 70}, 9 (1993);
  ``{\em Critical Behavior in Massless
Scalar Field Collapse}," in {\it Approaches to Numerical
Relativity, Proceedings of the International Workshop on Numerical
Relativity}, Southampton, December, 1991, Edited by Ray d'Inverno;
``{\em Critical Behavior in Scalar Field Collapse},"
in {\it Deterministic Chaos in General Relativity}, Edited by D.
Hobill et al. (Plenum Press, New York, 1994), pp. 155-175.

\bibitem{Golden} G.I. Barenblatt, {\em Similarity, Self-Similarity, and
Intermediate Asymptotics} (Consultants Bureau, New York, 1979); N.
Goldenfeld, {\em Lectures on Phase Transitions and the
Renormalization Group} (Addison Wesley Publishing Company, New
York, 1992).

\bibitem{Gun00}  C. Gundlach,
``{\em Critical phenomena in gravitational collapse: Living
Reviews}," {\tt gr-qc/0001046} (2000), and references therein.

\bibitem{Wang01} A.Z. Wang, ``{\em Critical Phenomena in Gravitational
Collapse: The Studies So Far}," {\tt gr-qc/0104073},  Braz. J.
Phys. {\bf 31}, 188 (2001), and references therein.

\bibitem{Cho03} M.W. Choptuik, E.W. Hirschmann, S.L. Liebling, and F.
Pretorius, Phys. Rev. {\bf D68}, 044007 (2003).

\bibitem{Wang03} A.Z. Wang,  Phys. Rev. {\bf D68}, 064006 (2003).

\bibitem{Even}   C. R. Evans and J. S. Coleman,
Phys. Rev. Lett. {\bf 72},
1782 (1994);  T. Koike, T. Hara, and S. Adachi, {\em ibid.}, {\bf
74}, 5170 (1995); C. Gundlach, {\em ibid.},   {\bf 75}, 3214
(1995); E. W. Hirschmann and D. M. Eardley,  Phys. Rev. {\bf D52},
5850 (1995).

\bibitem{Brady02} P. R. Brady, M. W. Choptuik, C. Gundlach, and
D. W. Neilsen, Class. Quantum Grav. {\bf 19},  6359 (2002).


\bibitem{PC00}  F. Pretorius and M. W. Choptuik,  Phys. Rev. {\bf D62},
124012 (2000).

\bibitem{HO01} V. Husain and M. Olivier, Class. Quantum
Grav. {\bf 18}, L1 (2001).

\bibitem {Gar01} D. Garfinkle, Phys. Rev. {\bf D63}, 044007
(2001).

\bibitem{CF01}  G. Cl\'ement and A. Fabbri, Class. Quantum Grav. {\bf
18}, 3665 (2001); Nucl. Phys. {\bf B630}, 269 (2002).

\bibitem{GG02} D. Garfinkle and C Gundlach, Phys. Rev. {\bf D66}, 044015
 (2002).

\bibitem{HWW04} E.W. Hirschmann, A.Z. Wang, and Y. Wu,
Class. Quant. Grav. {\bf 21}, 1791 (2004).

\bibitem {Chandra83}  S. Chandrasekhar, {\em The Mathematical Theory
of Black Holes} (Clarendon Press, Oxford University Press, Oxford,
1983).

\bibitem{CH89} B. Carter and R.N. Henriksen, Ann.
Physique Suppl. {\bf 14},
47 (1989). See also, A.A. Coley, Class. Quantum Grav. {\bf 14}, 87
(1997).

\bibitem{CT71} M.E. Cahill and A.H. Taub, Commun. Math.
Phys. {\bf 21}, 1 (1971).

\bibitem{CFJW04} R. Chan, M.F.A. da Silva, J.F. Villas da Rocha, A. Wang, 
Int. J. Mod. Phys. D, in press, {\tt gr-qc/0406026} (2005).

\bibitem{Yasuda} A. Y. Miguelote, N. A. Tomimura, and A.Z. Wang, Gen. Rel.
Grav. {\bf 36}, 1883 (2004).

\bibitem{Kramer80}  D. Kramer, H. Stephani, E. Herlt, and M.
MacCallum, {\em Exact Solutions of Einstein's Field Equations}
(Cambridge University Press, Cambridge, England, 1980).

\bibitem{VS} A. Vilenkin and E. P. S. Shellard, {\it Cosmic String and Other
Topological Defects} (Cambridge University Press, Cambridge, 1994).



\end{thebibliography}
\end{document}